\documentclass[twocolumn]{aastex631}

\shorttitle{A Rosetta Stone for eccentricity}
\shortauthors{Knee et al.}
\graphicspath{{./}{figures/}}

\begin{document}

\title{A Rosetta Stone for eccentric gravitational waveform models}

\correspondingauthor{Alan M. Knee}
\email{aknee@phas.ubc.ca}

\author[0000-0003-0703-947X]{Alan M. Knee}
\affiliation{Department of Physics \& Astronomy, University of British Columbia, Vancouver, BC V6T 1Z1, Canada}

\author[0000-0002-4181-8090]{Isobel M. Romero-Shaw}
\affiliation{School of Physics \& Astronomy, Monash University, Clayton, VIC 3800, Australia}
\affiliation{OzGrav: The ARC Centre of Excellence for Gravitational Wave Discovery, Clayton, VIC 3800, Australia}
\affiliation{Department of Applied Mathematics and Theoretical Physics, Cambridge CB3 0WA, United Kingdom}

\author[0000-0003-3763-1386]{Paul D. Lasky}
\affiliation{School of Physics \& Astronomy, Monash University, Clayton, VIC 3800, Australia}
\affiliation{OzGrav: The ARC Centre of Excellence for Gravitational Wave Discovery, Clayton, VIC 3800, Australia}

\author[0000-0003-0316-1355]{Jess McIver}
\affiliation{Department of Physics \& Astronomy, University of British Columbia, Vancouver, BC V6T 1Z1, Canada}

\author[0000-0002-4418-3895]{Eric Thrane}
\affiliation{School of Physics \& Astronomy, Monash University, Clayton, VIC 3800, Australia}
\affiliation{OzGrav: The ARC Centre of Excellence for Gravitational Wave Discovery, Clayton, VIC 3800, Australia}



\begin{abstract}

Orbital eccentricity is a key signature of dynamical binary black hole formation. The gravitational waves from a coalescing binary contain information about its orbital eccentricity, which may be measured if the binary retains sufficient eccentricity near merger. Dedicated waveforms are required to measure eccentricity. Several models have been put forward, and show good agreement with numerical relativity at the level of a few percent or better. However, there are multiple ways to define eccentricity for inspiralling systems, and different models internally use different definitions of eccentricity, making it difficult to directly compare eccentricity measurements. In this work, we systematically compare two eccentric waveform models, \texttt{SEOBNRE} and \texttt{TEOBResumS}, by developing a framework to translate between different definitions of eccentricity. This mapping is constructed by minimizing the relative mismatch between the two models over eccentricity and reference frequency, before evolving the eccentricity of one model to the same reference frequency as the other model. We show that for a given value of eccentricity passed to \texttt{SEOBNRE}, one must input a $20$-$50\%$ smaller value of eccentricity to \texttt{TEOBResumS} in order to obtain a waveform with the same empirical eccentricity. We verify this mapping by repeating our analysis for eccentric numerical relativity simulations, demonstrating that \texttt{TEOBResumS} reports a correspondingly smaller value of eccentricity than \texttt{SEOBNRE}.

\end{abstract}



\section{Introduction} \label{sec:intro}


The LIGO-Virgo-KAGRA (LVK) collaboration has announced a total of 90 likely detections\footnote{This is the total number of candidate signals with at least a 50\% probability of being astrophysical. Different thresholds can yield different numbers of detections.} of gravitational waves (GWs) from the first three observing runs of the Advanced LIGO \citep{2015CQGra..32g4001L}, Advanced Virgo \citep{2015CQGra..32b4001A}, and KAGRA \citep{KAGRA:2018plz} interferometers. These detections are consistent with the inspiral and subsequent merger and ringdown of coalescing compact binaries, including binary black hole (BBH), binary neutron star, and neutron star-black hole systems \citep{PhysRevX.9.031040, LIGOScientific:2020ibl, LIGOScientific:2021usb, 2021arXiv211103606T}. Analyses of the publicly available strain data have yielded additional likely detections \citep{Nitz:2021uxj, Olsen:2022pin, Venumadhav:2019lyq}. BBH mergers constitute the majority of the detections thus far. By studying their GW emissions, the source properties of BBH systems can be inferred \citep[e.g.,][]{LIGOScientific:2016vlm, LIGOScientific:2020ufj}, including the masses of the two black hole companions, their spins, and their orbital eccentricities. 

Precise measurements of the intrinsic parameters describing BBH systems can reveal important clues about how such systems are formed \citep{LIGOScientific:2018jsj, 2021ApJ...913L...7A, LIGOScientific:2021psn}. In general, coalescing BBH systems can be formed via one of two proposed formation channels: isolated binary evolution, facilitated by, e.g., common envelope \citep{1998ApJ...506..780B, Belczynski_2002, 10.1046/j.1365-8711.2003.06616.x} or chemically homogeneous evolution \citep{2016MNRAS.458.2634M, 2016MNRAS.460.3545D}, and dynamical assembly within dense stellar environments, such as in the cores of globular \citep{1993Natur.364..423S, PortegiesZwart:1999nm, Portegies_Zwart_2002, Rodriguez:2016kxx} and nuclear \citep{Belczynski:2020bnq, Gerosa:2021mno} star clusters. 

Orbital eccentricity is a promising signature of dynamical assembly. Gravitational radiation efficiently circularizes binary systems \citep{PhysRev.131.435, PhysRev.136.B1224}, and any initial eccentricity present at the epoch of isolated binary formation is expected to be almost completely damped away when the GW signal reaches the observing band of ground-based detectors ($\sim 10$ Hz). Dynamical binaries, however, can sometimes form with sufficiently small separation such that there is insufficient time for the system to fully circularize before its radiation becomes observable, allowing these sources to be distinguished from isolated binaries based on their eccentricity \citep{Rodriguez:2018pss, Rodriguez:2017pec, Lower:2018seu, Samsing:2017xmd, Zevin:2018kzq, Zevin:2021rtf}. Possible hints of eccentricity have already been detected in a few BBH candidates observed by the LVK network \citep{Romero-Shaw:2019itr, Romero-Shaw:2021ual, Romero-Shaw:2022xko}, with GW190521 being a notable case study \citep{Romero-Shaw:2020thy, Gayathri:2020coq}. As the number of detections increases, measurements of eccentricity will place tighter constraints on the fraction of BBH mergers which are of dynamical origin.

Currently, the literature features two eccentric inspiral-merger-ringdown waveform models that have been used for some form of GW parameter estimation: \texttt{TEOBResumS} \citep{Damour:2014sva, Nagar:2015xqa, Nagar:2018zoe, Nagar:2019wds, Nagar:2020pcj, Riemenschneider:2021ppj, Chiaramello:2020ehz, Nagar:2021gss} and \texttt{SEOBNRE} \citep{Cao:2017ndf, Liu:2019jpg, Liu:2021pkr}.\footnote{\texttt{ENIGMA} \citep{Huerta:2017kez} is another eccentric model, but has so far only been used for creating injections that were later recovered with a quasi-circular approximant. There are also the inspiral-only models \texttt{EccentricFD} \citep{PhysRevD.90.084016} and \texttt{TaylorF2e} \citep{Moore:2019xkm}, which were used by \citet{Wu:2020zwr} and \citet{Lenon:2020oza} to analyze BBH and binary neutron star events, respectively. For high-mass eccentric candidates like GW190521, it is essential to use waveforms that include merger physics.} These are time-domain, aligned-spin models which employ the effective one-body (EOB) formalism \citep{PhysRevD.59.084006, PhysRevD.62.064015} to solve the general-relativistic two-body problem. Both waveforms have been validated against numerical relativity simulations of eccentric BBH systems \citep{Hinder:2017sxy} with low-to-moderate eccentricity ($\lesssim 0.3$), giving mismatch factors (defined later in Equation \ref{eq:mismatch}) no larger than 2-3\% \citep{Cao:2017ndf, Chiaramello:2020ehz}. In addition to \texttt{TEOBResumS} and \texttt{SEOBNRE}, several other eccentric waveform models are under development \citep[e.g.,][]{Ramos-Buades:2021adz, Islam:2021mha, Chen:2020lzc, PhysRevD.103.124053, PhysRevD.103.124011, Hinder:2017sxy}.


In parameter estimation, the waveform model typically needs to be evaluated on the order of $10^5$ times or more. Neither of the waveforms introduced above are fast enough for direct use in parameter estimation in their current ``out-of-the-box'' states, and certain workarounds are required. 
\citet{Romero-Shaw:2019itr, Romero-Shaw:2020thy, Romero-Shaw:2021ual, Romero-Shaw:2022xko} constrained the eccentricities of 62 BBH candidates with \texttt{SEOBNRE} by importance-sampling the parameter space with a computationally cheaper quasi-circular waveform, and subsequently reweighting to the eccentric posterior \citep[e.g.,][]{Payne:2019wmy}.
Another study by \citet{OShea:2021ugg} reanalyzed two BBH candidates with \texttt{TEOBResumS} by loosening the error tolerance of the ODE integrator to speed up evaluation times. More recently, \citet{Iglesias2022} reanalyzed five BBH candidates with \texttt{TEOBResumS} and the rapid parameter estimation algorithm \texttt{RIFT} \citep{Lange:2018pyp}. Though they did not conduct parameter estimation, \citet{Zevin:2021rtf} also used \texttt{TEOBResumS} to evaluate the impact of selection effects on the detection of eccentric BBH mergers.

By analyzing individual GW events with different waveform models, we can learn more about the physics of the source system, as well as build confidence in our inferences if the models yield similar results. However, the notion of eccentricity is inherently ambiguous in general relativity, where particles in a two-body system are not restricted to perfectly elliptical orbits. In practice, different eccentric waveform models apply different definitions of eccentricity, and so it is currently not straightforward to reconcile measurements obtained with different eccentric waveforms.

As interest in eccentric binaries grows, it is becoming increasingly important for our analyses to implement consistent definitions of eccentricity. In this work, we address this problem by constructing a mapping between the eccentricity definitions of \texttt{TEOBResumS} and \texttt{SEOBNRE}. Our approach is to minimize the relative mismatch between these two models over the parameters of interest, namely their eccentricity and reference frequency. We then evolve the eccentricities to the same frequency. Using this mapping, a measurement of eccentricity obtained with one model can be converted into an equivalent eccentricity measured by the other model, allowing for a more direct comparison. 

The rest of this paper is organized as follows: In Section \ref{sec:bg}, we discuss the challenges of defining eccentricity in general relativity and how this leads to discrepancies between waveform models; in Section \ref{sec:method}, we outline our method for mapping the eccentricities defined by \texttt{TEOBResumS} and \texttt{SEOBNRE}; in Section \ref{sec:results} we discuss the results of our eccentricity mapping analysis; finally, in Section \ref{sec:summary}, we offer concluding remarks.

\section{Background} \label{sec:bg}

\subsection{Eccentric compact binaries} \label{sec:eccbin}

Eccentricity is most intuitively understood in the Newtonian regime, where it simply describes the amount by which an orbit deviates from a circle. In the Keplerian parameterization, the eccentricity, $e$, is defined through the orbital equation of motion
\begin{equation} \label{eq:keplecc}
    r(t) = \frac{a(1-e^2)}{1+e\cos\phi(t)}\,,
\end{equation}
where $r$ is the radial distance from the focus, $a$ is the semimajor axis, and $\phi$ is the true anomaly. An eccentricity of $e=0$ is a circular orbit, $0<e<1$ is an eccentric orbit, and $e\geq1$ is a parabolic (if equal) or hyperbolic unbound orbit. For a binary system with component masses $m_{1,2}$, each with semimajor axes $a_{1,2}$, the orbital (angular) frequency obeys Kepler's third law,
\begin{equation} \label{eq:kepler3}
    \bar{\omega}^2a^3=GM\,,
\end{equation}
where $M=m_1+m_2$ is the total mass, and $a=a_1+a_2$ is the sum of the semimajor axes of each object. The quantity $\bar{\omega}$ represents an average orbital frequency, which we refer to as the ``Keplerian'' frequency. Ignoring the inspiral for a moment and considering just the motion of the two objects along unchanging, closed orbits, the {\it instantaneous} orbital frequency is given by \citep[e.g.,][]{2014grav.book.....P}
\begin{equation} \label{eq:omegatN}
    \omega(t)=\frac{\bar{\omega}\sqrt{1-e^2}}{[1-e\cos E(t)]^2}
    \approx \bar{\omega}[1+2e\cos(\bar{\omega}t)]\,,
\end{equation}
where $E(t)$ is the eccentric anomaly, and the second expression follows in the low-eccentricity limit. Thus, in eccentric binaries the frequency consists of a Keplerian component, $\bar{\omega}$, which is constant in the absence of radiation-reaction, plus an oscillating component with amplitude $2e\bar{\omega}$ and period equal to that of the orbit, which vanishes in the quasi-circular limit. The same is true of the GW signal, which in the case of eccentric systems receives contributions from several harmonics of the orbital frequency, instead of the power being concentrated in the second harmonic as in quasi-circular systems \citep{PhysRev.131.435}. The emission of GWs is also asymmetric, with greater power radiated during a periastron passage compared to apastron.

This picture becomes more complicated when we properly account for the energy lost in GWs, introducing a dissipative radiation-reaction force which causes the system to inspiral. At leading order, the semimajor axis and eccentricity of the binary will decay according to Peters' equations \citep{PhysRev.136.B1224}:
\begin{equation} \label{eq:dadt}
    \bigg\langle\frac{{\rm d}a}{{\rm d}t}\bigg\rangle = -\frac{64}{5}\frac{G^3m_1m_2M}{c^5a^3(1-e^2)^{7/2}}\bigg(1+\frac{73}{24}e^2+\frac{37}{96}e^4\bigg)\,;
\end{equation}
\begin{equation} \label{eq:dedt}
    \bigg\langle\frac{{\rm d}e}{{\rm d}t}\bigg\rangle = -\frac{304}{15}\frac{G^3m_1m_2M}{c^5a^4(1-e^2)^{5/2}}\bigg(1+\frac{121}{304}e^2\bigg)\,,
\end{equation}
where angle brackets indicate that the derivatives are {\it orbit-averaged} quantities. It is important to note that the Keplerian interpretation of eccentricity is only meaningful in an adiabatic sense, when the system is evolving slowly enough such that the orbits remain nearly closed on orbital timescales. The adiabatic approximation is thus accurate during the early inspiral, but inevitably breaks down closer to merger. In the highly relativistic regime, alternative ways of measuring eccentricity which better reflect the true dynamics of the system are needed. 

Eccentricity is challenging to define rigorously in general relativity. At the core of this problem is the fact that, while its physical effect on GW emission is observable, eccentricity is nonetheless gauge-dependent. As a result, there is no single definition of eccentricity in general relativity that can be applied universally, and several definitions have been conceived within different contexts \citep[see e.g.,][for a review]{Loutrel:2018ydu}.
In the post-Newtonian formalism, eccentricity is frequently expressed in terms of the orbit-averaged quantities $e_r$, $e_t$, and $e_\phi$, which is known as the quasi-Keplerian parameterization \citep{AIHPA_1985__43_1_107_0, AIHPA_1986__44_3_263_0, Blanchet:2013haa}. In contrast, the so-called osculating method for measuring the eccentricity \citep{Damour:2004bz, Pound:2010pj} does not use orbit-averaged quantities, and yields behaviour that apparently contradicts post-Newtonian theory \citep{Loutrel:2018ssg}. In the field of numerical relativity, eccentricity is usually measured in terms of the coordinate separation between the binary components \citep{Boyle:2007ft, Pfeiffer:2007yz, Tichy:2010qa}, or the orbital or GW frequency of the system \citep{Mroue:2010re, Buonanno:2010yk, Ramos-Buades:2018azo}, neither of which provide exactly the same measurement since these methods assume a coordinate system and are therefore gauge-dependent.

\subsection{Eccentric waveforms} \label{sec:eccwave}

In waveform modelling, it is typical to define time-dependent quantities such as eccentricity, inclination angle and spin tilts in terms of a reference frequency.\footnote{Recent work, e.g.,  by \cite{Mould2022}, seeks to  define time-dependent quantities at past time infinity.}
Eccentric models therefore utilize a reference eccentricity, $e_0$, defined at some GW reference frequency, $f_{\rm ref}$, which are both taken as inputs by the models, in addition to the masses and spins. Note that these models do not employ any cosmology, and therefore the parameters are implicitly defined in the detector frame. For \texttt{TEOBResumS} and \texttt{SEOBNRE}, the input eccentricity and reference frequency are used to determine a set of adiabatic initial conditions from which the trajectories of the binary components are evolved. It is important to emphasize that the eccentricities supplied to the waveform models do not correspond to systems with the same physical eccentricity, and therefore these ``waveform eccentricities'' carry different meaning depending on the waveform model. 

\begin{figure}[t!]
    \epsscale{1.15}
    \plotone{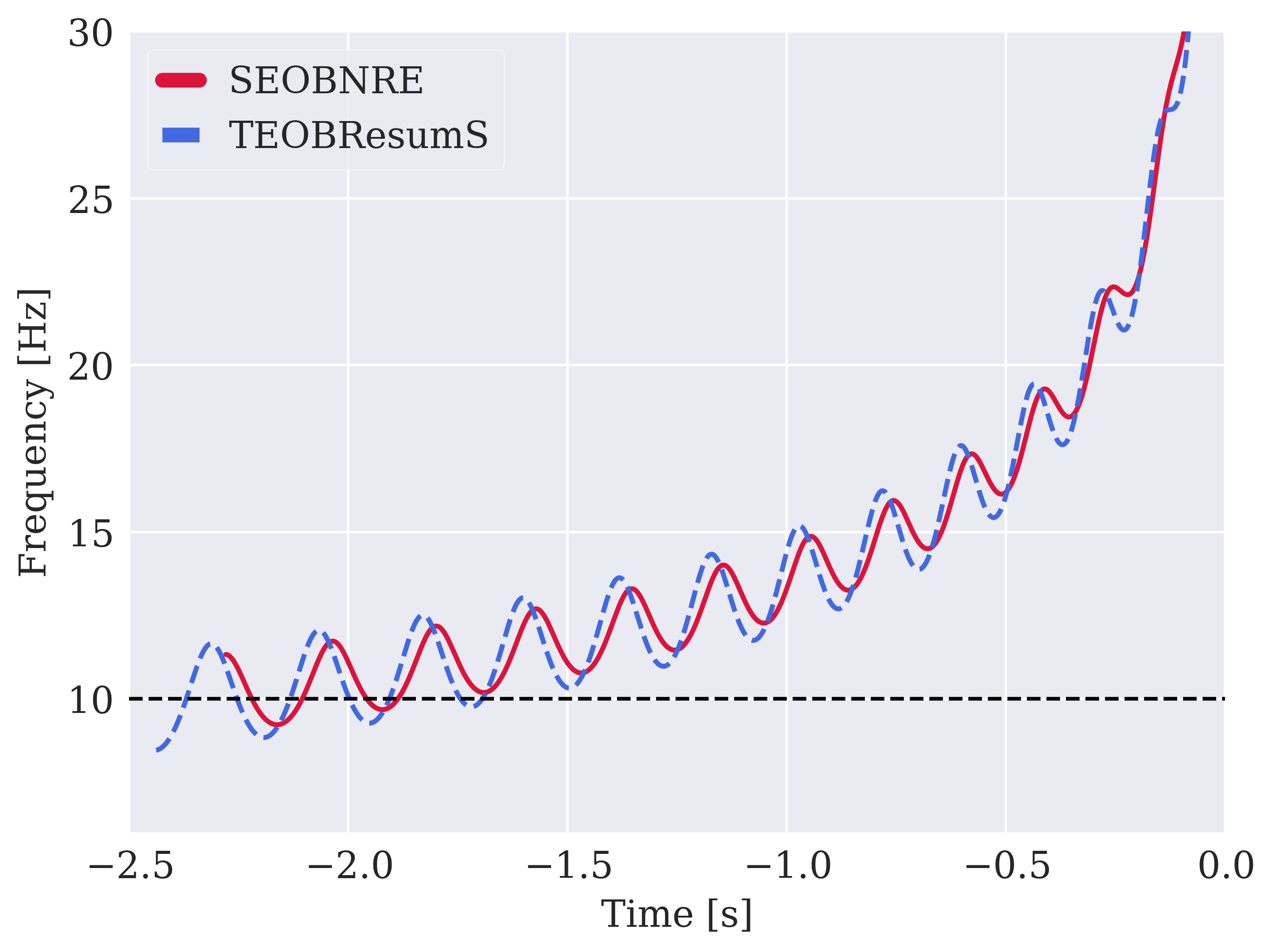}
    \caption{Comparison of the two waveform models, \texttt{TEOBResumS} and \texttt{SEOBNRE}, for an equal mass, non-spinning BBH with total mass $M_{\rm det}=100\,M_\odot$ in the detector frame, showing the $(\ell=2, |m|=2)$ waveform frequency as a function of time. The waveforms are given identical input parameters, with waveform eccentricity $e_0=0.1$ at a reference frequency of $10$~Hz (indicated by a horizontal line). For the same input value of eccentricity, the frequency oscillations in \texttt{TEOBResumS} are noticeably larger, implying a higher simulated eccentricity. \texttt{TEOBResumS} also interprets the reference frequency differently (described in Section \ref{sec:eccwave}), causing it to start at a lower frequency than \texttt{SEOBNRE}. Lastly, the waveforms start from different points in the orbit: \texttt{TEOBResumS} starts from apastron (a trough), and \texttt{SEOBNRE} starts from periastron (a peak). \label{fig:comparison}}
\end{figure}

The instantaneous GW frequency calculated with \texttt{TEOBResumS} and \texttt{SEOBNRE} (using identical parameters) are compared in Fig.~\ref{fig:comparison}. For non-precessing binaries, the $(\ell=2, |m|=2)$ waveform can be simply decomposed into an amplitude and complex phase,
\begin{equation} \label{eq:decomp}
    h_+(t)-ih_\times(t) = A_{22}(t)\exp[-i\Phi_{22}(t)]\,,
\end{equation}
where $h_{+,\times}$ are the two GW polarizations, and the GW frequency is given by the time derivative of the phase,
\begin{equation}
    2\pi f_{22}(t) = \frac{{\rm d}\Phi_{22}}{{\rm d}t}\,.
\end{equation}
The shapes of the orbits are encoded in the frequency evolution. As explained in Section \ref{sec:eccbin}, the frequency consists of a monotonically increasing Keplerian component, and an eccentricity-induced oscillating component. 
The amplitude of this oscillating component provides a direct measure of the eccentricity of the system, and can be used to compare the eccentricity of two different waveform families.
Furthermore, we can identify local maxima in the frequency oscillations with periastron passages, and local minima with apastron passages. 
The oscillations become suppressed as the inspiral progresses due to the circularization of the system.

The eccentric EOB models we examine in this work, \texttt{TEOBResumS} and \texttt{SEOBNRE}, ostensibly use the Keplerian definition of eccentricity, yet produce waveform templates with apparently different eccentricities as determined by the amplitude of the frequency oscillations, from which one can derive an empirical measure of eccentricity \citep{Mora:2002gf}
\begin{equation} \label{eq:eomega}
    e_\omega = \frac{\omega_{\rm p}^{1/2}-\omega_{\rm a}^{1/2}}{\omega_{\rm p}^{1/2}+\omega_{\rm a}^{1/2}}\,,
\end{equation}
where $\omega_{\rm p,a}$ are the orbital or GW frequencies measured at periastron and apastron passage, respectively.
There are multiple reasons for this discrepant behaviour:
\begin{enumerate}
  \item 
  There are a number of differences between the two waveform models related to their treatment of the conservative dynamics of the system and the radiation-reaction under eccentric conditions. Furthermore, the models employ a different logic for setting the initial conditions of the system, which couples with the above factors to render an effectively model-dependent definition of eccentricity. Given the same input value of eccentricity, the waveforms end up simulating different levels of eccentricity, as judged by Equation \ref{eq:eomega}.
  \item The models are calibrated differently to numerical relativity simulations, which can yield additional mismatch at merger.
  \item 
  The models use different conventions for the reference frequency, $f_{\rm ref}$, which the waveforms also use as the starting frequency of the system. \texttt{SEOBNRE} defines $f_{\rm ref}$ with respect to the Keplerian frequency of the GW radiation. For \texttt{TEOBResumS}, the user is able to define $f_{\rm ref}$ as either the GW frequency at periastron, apastron, or the mean of the two. The last option is closest to what \texttt{SEOBNRE} uses, but it is still not equivalent. For the same value of $f_{\rm ref}$, \texttt{TEOBResumS} in general begins from a lower initial frequency than \texttt{SEOBNRE}, since its definition corresponds to a slightly smaller Keplerian frequency than \texttt{SEOBNRE}. For waveform eccentricity $e_0=0.2$, the difference in starting frequency reaches about 10\% which, at lower masses, translates into substantial difference in waveform length.\footnote{For example, changing the initial frequency of a $250\,M_\odot$ system from $10$~Hz to $9$~Hz increases the time to merger by $0.15$~seconds, but increases it by $2.7$~seconds for a $50\,M_\odot$ system.} This means that in addition to the models using intrinsically different eccentricity definitions, \texttt{TEOBResumS} parameterizes eccentricity at an earlier reference point than \texttt{SEOBNRE}.
  \item Neither model admits the mean anomaly as an input parameter. Instead, the models initialize the system from opposite orbital configurations: \texttt{TEOBResumS} always starts from apastron, \texttt{SEOBNRE} always starts from periastron. Because of this, the arguments of periastra are often misaligned unless one carefully tunes the reference frequency to avoid a persistent phase offset in the frequency oscillations. Work is underway to incorporate a variable mean anomaly in parameter estimation with these models \citep{Islam:2021mha}; however, at current detector sensitivity, neglecting this parameter is not expected to bias results \citep{Clarke2022}.
\end{enumerate}
Some or all of these factors may be present when comparing other eccentric waveform models.

\begin{figure}[t!]
    \epsscale{1.15}
    \plotone{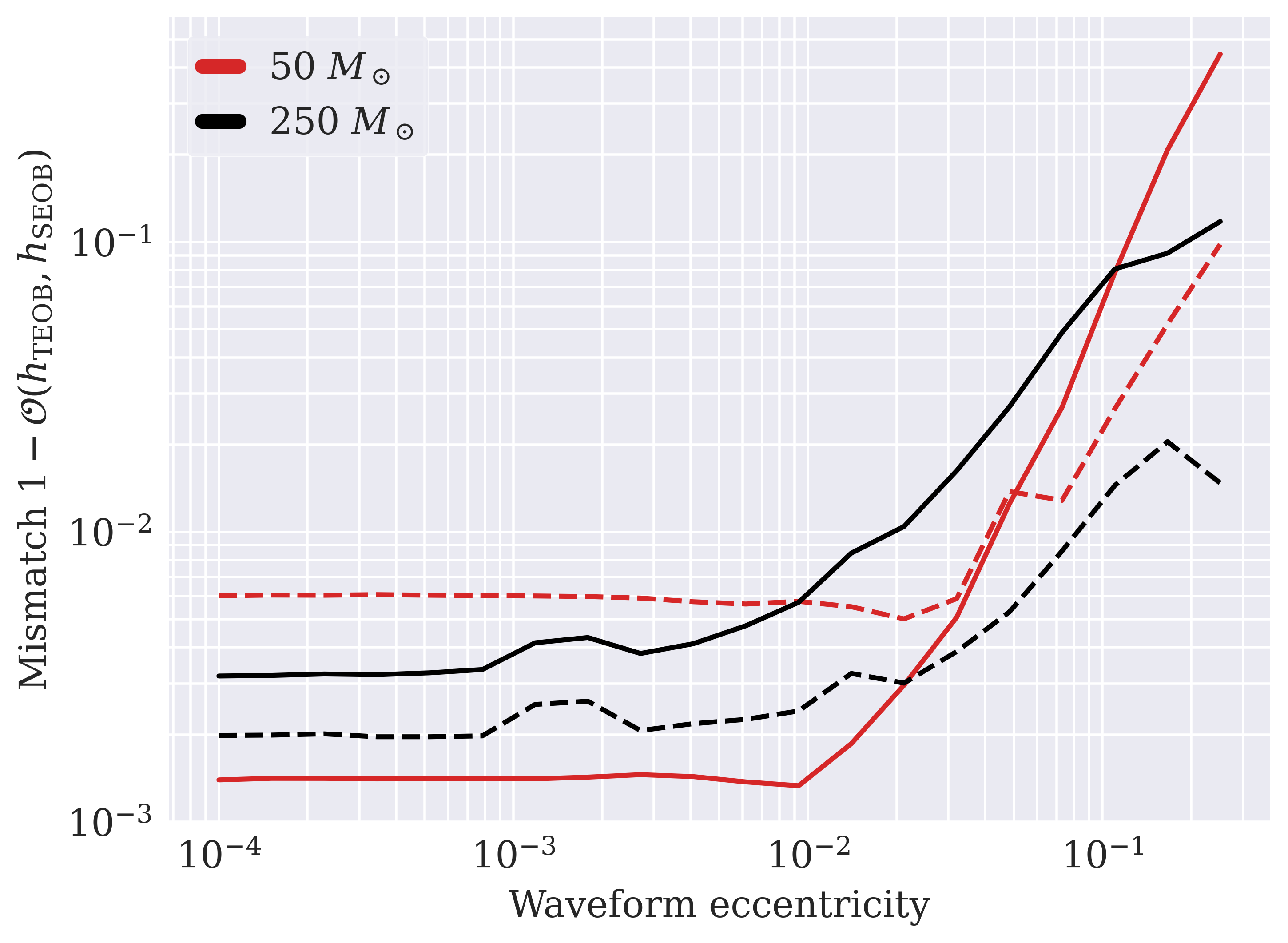}
    \caption{Mismatch between \texttt{TEOBResumS} and \texttt{SEOBNRE} as a function of waveform eccentricity at $10$~Hz, showing how the mismatch increases with higher eccentricity. Curves are shown for equal mass systems with different total masses as indicated. The reference frequencies supplied to \texttt{TEOBResumS} are adjusted so that the waveform starts with a Keplerian frequency of $10$~Hz, ensuring consistency with the way \texttt{SEOBNRE} defines the reference frequency. We show mismatches calculated for both white noise (solid curves) and LVK sensitivity (dashed curves). \label{fig:mismatch}}
\end{figure}

\begin{figure}[t!]
    \epsscale{1.15}
    \plotone{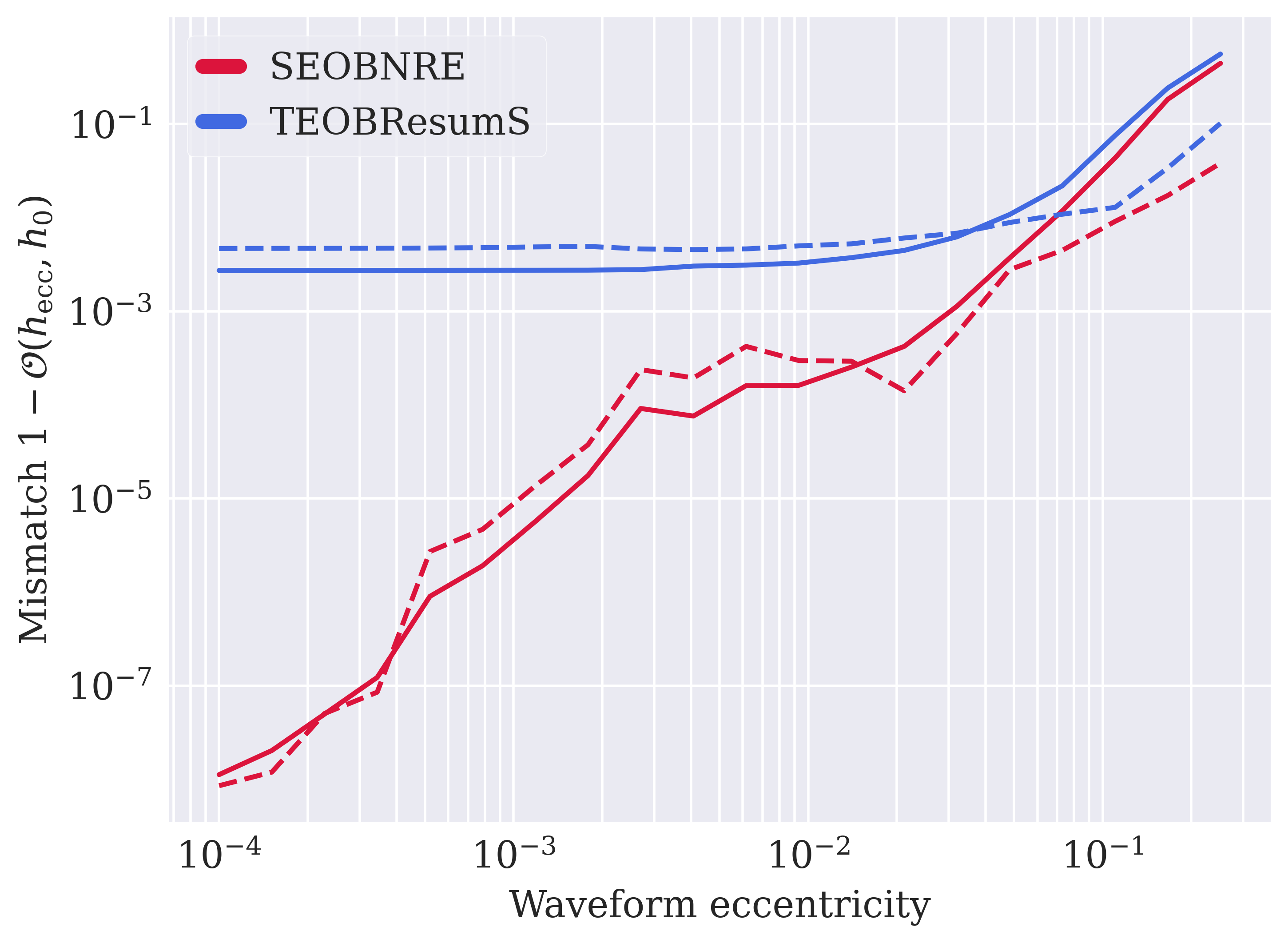}
    \caption{Mismatch between the eccentric and quasi-circular version of \texttt{TEOBResumS} and \texttt{SEOBNRE} ($50\,M_\odot+50\,M_\odot$, non-spinning), plotted as a function of waveform eccentricity at $10$~Hz. As the eccentricity of \texttt{SEOBNRE} decreases, there is a corresponding decrease in mismatch with its quasi-circular counterpart. In contrast, the mismatch between the eccentric model \texttt{TEOBResumS-DALI} and the quasi-circular model \texttt{TEOBResumS-GIOTTO} levels off at a minimum value, reflecting a discontinuity between the behaviours of the two models. As in Figure \ref{fig:mismatch}, the mismatches are shown for white noise (solid curves) and LVK sensitivity (dashed curves), and we use the same frequency adjustment for \texttt{TEOBResumS}. \label{fig:mismatch_circ}}
\end{figure}

The amount of agreement between two complex waveforms, $h_1$ and $h_2$, is quantified with their match (or overlap), defined through a noise-weighted inner product in some frequency band maximized over a time and phase of coalescence \citep{Flanagan:1997kp, Lindblom:2008cm},
\begin{equation}\label{eq:mismatch}
    \mathcal{O}(h_1,h_2) = \max_{t_0,\Phi_0} \frac{\langle h_1|h_2 \rangle}{\sqrt{\langle h_1|h_1 \rangle \langle h_2|h_2 \rangle}}\,,
\end{equation}
where
\begin{equation}
    \langle h_1|h_2 \rangle = 4\Re\int_{f_{\rm low}}^{f_{\rm high}} \frac{\tilde{h}_1(f)\tilde{h}_2^*(f)}{S_{\rm n}(f)}\,{\rm d}f\,.
\end{equation}
Here, tildes denote a Fourier transform, and $S_{\rm n}(f)$ is the noise power spectral density (PSD) in the detector. The match is normalized such that a value of $\mathcal{O}=1$ means the waveforms are identical (within the specified band), and $\mathcal{O}=0$ means the waveforms are maximally different. The difference between the waveforms is thus given by their mismatch, $1-\mathcal{O}(h_1,h_2)$. In Figure \ref{fig:mismatch}, we show the relative mismatch between \texttt{TEOBResumS} and \texttt{SEOBNRE} as a function of waveform eccentricity. To make the comparison clearer, we use Equation \ref{eq:omegatN} to adjust the reference frequency\footnote{This formula is only an approximation because the eccentricities are not interchangeable. However, we find that it works well within the eccentricity range explored by this work.} supplied to \texttt{TEOBResumS}, ensuring that it is initialized with the same initial Keplerian frequency as \texttt{SEOBNRE}. If we do not do this, the mismatch oscillates as the eccentricity changes due to the arguments of periastra going in and out of phase. With consistent reference frequencies, we can more clearly observe that the mismatch between the models increases as the eccentricity increases, reaching $1\%$ for $e_0\sim 0.05$ and exceeding $10\%$ for $e_0\sim 0.2$ (with white noise). The models also differ in their handling of the transition from eccentric to quasi-circular dynamics, which we show in Figure \ref{fig:mismatch_circ} by computing the mismatch between the eccentric ($e_0>0$) and quasi-circular ($e_0=0$) configurations of each model as a function of waveform eccentricity. Figure \ref{fig:mismatch_circ} shows that \texttt{SEOBNRE} approaches quasi-circular behaviour as $e_0\rightarrow0$. In contrast, the eccentric configuration of \texttt{TEOBResumS}, known as \texttt{TEOBResumS-DALI}, displays a minimum level of mismatch with its quasi-circular counterpart, \texttt{TEOBResumS-GIOTTO}, for arbitrarily small (but non-zero) eccentricities, reflecting the systematic differences between the models.\footnote{Specifically, \texttt{SEOBNRE} utilizes quasi-circular initial conditions \citep{Buonanno:2005xu} adjusted by an eccentric factor \citep{Cao:2017ndf}, and so it obtains exactly quasi-circular initial conditions in the limit of zero eccentricity. \texttt{TEOBResumS-DALI} implements post-adiabatic initial conditions \citep{Chiaramello:2020ehz, Damour:2012ky, Hinderer:2017jcs}, but is discontinuous with \texttt{TEOBResumS-GIOTTO}, in part due to a difference in the description of radiation-reaction which does not vanish in the limit of $e_0\rightarrow 0$.}

\section{Method} \label{sec:method}

In order to construct a map between different definitions of eccentricity, we find the optimal values of the waveform eccentricity and reference frequency which minimize the mismatch (Equation \ref{eq:mismatch}), given fixed component masses and aligned spin. This will tell us what input eccentricity we need to pass to \texttt{TEOBResumS} to obtain a waveform template that is as similar as possible to a given \texttt{SEOBNRE} template, and vice versa. 

Our strategy is as follows. Since \texttt{SEOBNRE} takes much longer to evaluate than \texttt{TEOBResumS}, we generate a fiducial set of 550 \texttt{SEOBNRE} waveforms with detector-frame total masses ranging from $50\,M_\odot$ to $300\,M_\odot$, and eccentricities from $e_0^{\rm SEOB}=0$ to $0.4$ (uniformly) at $10$~Hz. We then minimize the mismatch with \texttt{TEOBResumS} by varying its waveform eccentricity and reference frequency,\footnote{We have to minimize over $f_{\rm ref}$ because the models are initialized from opposite starting points in the orbit. We adopt the Keplerian convention for the reference frequency, as explained in Section \ref{sec:eccwave}, meaning we need to use Equation \ref{eq:omegatN} to adjust the reference frequency passed to \texttt{TEOBResumS} so that it starts from the desired Keplerian frequency.} keeping all other parameters identical to \texttt{SEOBNRE}. To examine how the eccentricity conversion varies with the mass ratio and spins, we also generate \texttt{SEOBNRE} waveforms with mass ratios from $q=m_1/m_2=1$ to $4$, and with component spins\footnote{Note that \texttt{SEOBNRE} is not valid for large spins $|\chi_{1,2}|\geq 0.6$.} between $\chi_1=\chi_2=-0.5$ and $0.5$. We consider only the case where the spins are aligned with the orbital angular momentum of the system (i.e.~non-precessing), as neither model allows for misaligned spins. For our mismatch calculations, we integrate over a frequency band starting from $10$~Hz and ending at the Nyquist frequency of $2048$~Hz. As the \texttt{SEOBNRE} model is initialized from periastron, the initial (instantaneous) frequency of the waveform will generally be above the requested frequency of $10$~Hz. We avoid any issues that may arise from having the waveforms begin in-band by generating them from a lower frequency than $10$~Hz. We use Peters' equations (Equations \ref{eq:dadt}--\ref{eq:dedt}) to back-evolve the eccentricity as needed, relating the semimajor axis to the GW frequency with Kepler's third law, $a=[GM/(\pi f_{\rm GW})^2]^{1/3}$. We back-evolve to $5$~Hz in all cases except for the $50\,M_\odot$ waveforms, for which we use $7$~Hz. Similarly, we ensure that the \texttt{TEOBResumS} waveforms do not start in-band by varying the reference frequency between $5$--$10$~Hz (or $7$--$10$~Hz in the low-mass case). The start of the inspirals are tapered with a half-Tukey window rolled over 0.2 seconds to prevent spectral leakage from contaminating the Fourier transforms. Only the (2,2)-mode waveforms are considered in our analysis, since the publicly available version of \texttt{SEOBNRE} does not currently include higher-order modes.

In order to combine information from both GW polarizations, we calculate the mismatch by substituting the detector response \citep{Thorne1987},
\begin{equation} \label{eq:detresp}
    h(t) = F_+(\alpha,\delta,\psi)h_+(t) + F_\times(\alpha,\delta,\psi) h_\times(t)\,,
\end{equation}
directly into Equation \ref{eq:mismatch}, where $F_{+,\times}$ are the antenna patterns, which are functions of the sky location of the source (right ascension, $\alpha$, and declination, $\delta$), and the polarization angle, $\psi$. We set the sky location by choosing the optimal orientation relative to the Livingston detector at some arbitrary reference time, setting $\psi=0$ and assuming face-on inclination.

\begin{figure*}[t!]
    \epsscale{1.125}
    \plotone{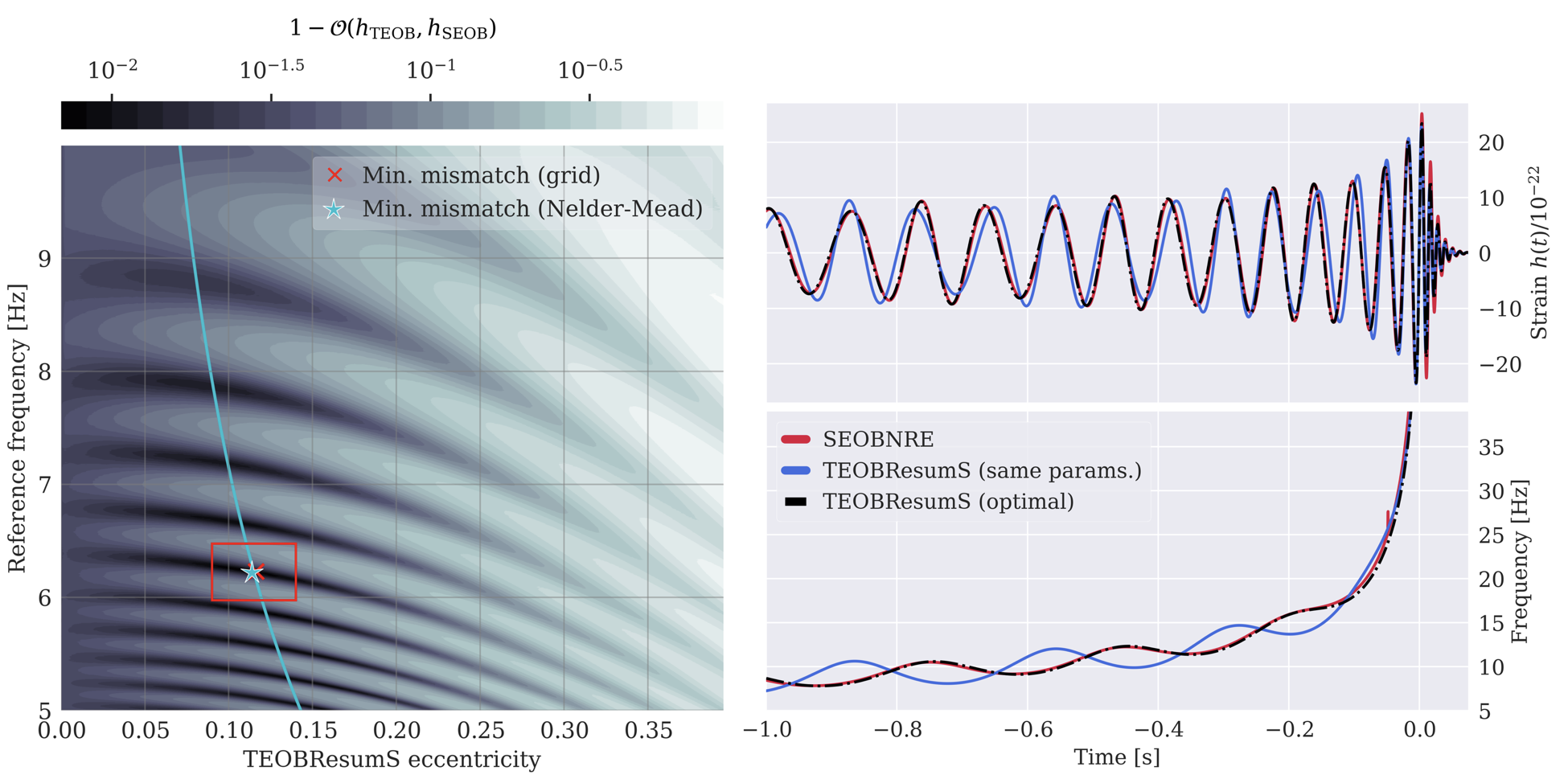}
    \caption{{\it Left panel}: Mismatch between \texttt{TEOBResumS} waveforms and a fiducial \texttt{SEOBNRE} waveform for an equal-mass non-spinning BBH with total mass $200\,M_\odot$ in the detector frame, assuming a white noise PSD. The \texttt{TEOBResumS} waveforms are distributed over a grid in waveform eccentricity and (Keplerian) reference frequency. The \texttt{SEOBNRE} waveform has $e_0^{\rm SEOB}=0.1$ at $10$~Hz. The colour scale shows the log-scaled mismatch between the waveforms at each grid point, where darker shading indicates a smaller mismatch. The regions of lower mismatch coincide with successive alignments of the arguments of periastron. The red cross marks the grid point with the lowest mismatch, which sits at the centre of a bounding box that is used as the search range for the optimization algorithm. The true minimum found by the optimizer is given by the cyan star, which is at a mismatch of $1-\mathcal{O}=0.0071$. For reference, the mismatch found when supplying identical parameters to both waveforms is $0.1567$. The cyan curve shows the eccentricity evolution track calculated with Peters' equations, which is used to forward-evolve the eccentricity to $10$~Hz. {\it Right panels}: Comparison of the fiducial \texttt{SEOBNRE} waveform (as observed in a detector; red curve) with two \texttt{TEOBResumS} waveforms, in which we supply either identical parameters as \texttt{SEOBNRE} (blue curve) or the optimal parameters (dashed black curve) from the parameter search that give the minimum mismatch. The top panel shows the time-domain strain, and the bottom panel shows the frequency evolution. The plotted waveforms are each aligned in coalescence phase and time. \label{fig:mismatch_grid}}
\end{figure*}

Due to the highly oscillatory behaviour of the mismatch as a function of eccentricity and reference frequency, we determine the minimum mismatch point using a two-step procedure. For each \texttt{SEOBNRE} template, we first perform a coarse grid search over the waveform eccentricity of the \texttt{TEOBResumS} model, $e_0^{\rm TEOB}$, and in the  reference frequency. Then, we obtain a more precise answer by using a Nelder-Mead optimizer to search over a small region centered on the lowest-mismatch grid point. An example mismatch grid is shown in the left panel of Figure \ref{fig:mismatch_grid}. By construction, the minimum mismatch point corresponds to an eccentricity measured at a lower frequency than $10$~Hz, where $e_{0}^{\rm SEOB}$ is defined. We correct for this by again using Peters' equations, this time to forward-evolve $e_0^{\rm TEOB}$ to $10$~Hz.\footnote{This introduces a small error on the evolved eccentricity, as Peters' equations are less accurate in this frequency range at higher masses.} 

We construct maps assuming both white noise (uniform) and LVK design-noise sensitivity PSDs. Using white noise tells us fundamentally how the eccentricity definitions differ, whereas using detector PSDs tells us what eccentricity one would actually measure with the two waveform models. These methods need not give the same answer. The detector curves weigh frequency bins differently, tending to weigh higher frequencies near merger more strongly at the expense of lower inspiral frequencies, where the eccentricity is larger. In contrast, white noise applies equal weighting at all frequencies. The results presented in Section \ref{sec:results} mainly use white noise, but we include examples of eccentricity maps computed with LVK noise in Appendix \ref{sec:appendix_lvk}.

\section{Results} \label{sec:results}

\subsection{Waveform eccentricity maps} \label{sec:maps}

\begin{figure*}
    \epsscale{1.05}
    \plotone{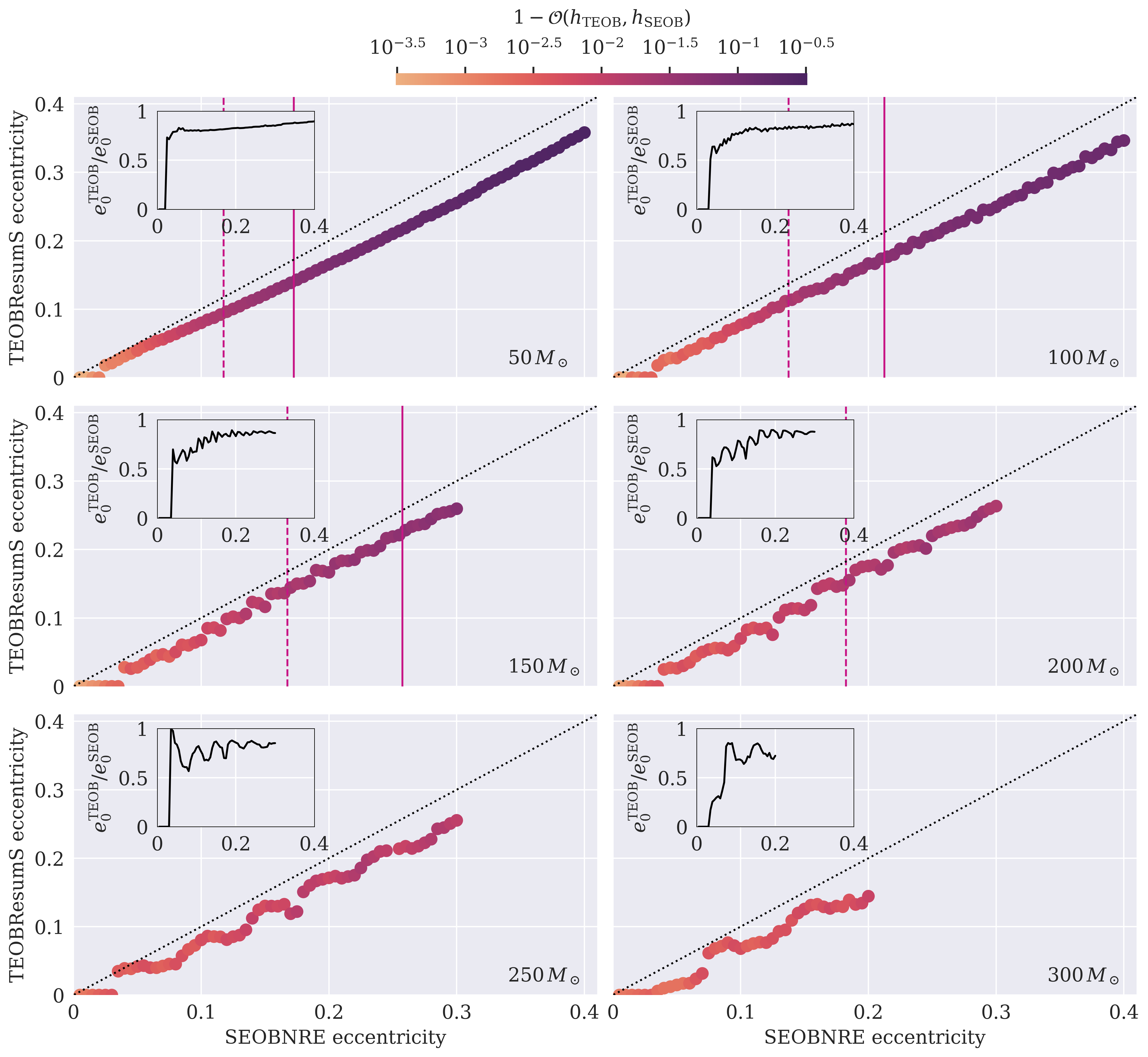}
    \caption{The mapping between the eccentricity definitions used by \texttt{SEOBNRE} and \texttt{TEOBResumS} obtained with our mismatch-minimization procedure, for various choices of total (detector-frame) mass, equal-masses and zero spins. The mismatch calculations assume white noise. All eccentricities are quoted at a reference GW frequency of $10$~Hz. The data points are coloured by the minimum mismatch between the \texttt{TEOBResumS} and \texttt{SEOBNRE} waveforms before evolving to $10$~Hz. The vertical lines indicate where the minimum mismatch crosses certain thresholds: waveforms to the right of the dashed line have $1-\mathcal{O}> 2\%$, and to the right of the solid line have $1-\mathcal{O}> 5\%$. The absence of any of these lines means the minimum mismatch never passes that threshold within the explored parameter ranges. The insets show the ratio of the two eccentricities as a function of $e_0^{\rm SEOB}$. To avoid unphysical model behaviour, we have to limit the maximum \texttt{SEOBNRE} eccentricity for heavier systems. \label{fig:ecc_conversion}}
\end{figure*}

\begin{figure*}
    \epsscale{1.01}
    \plotone{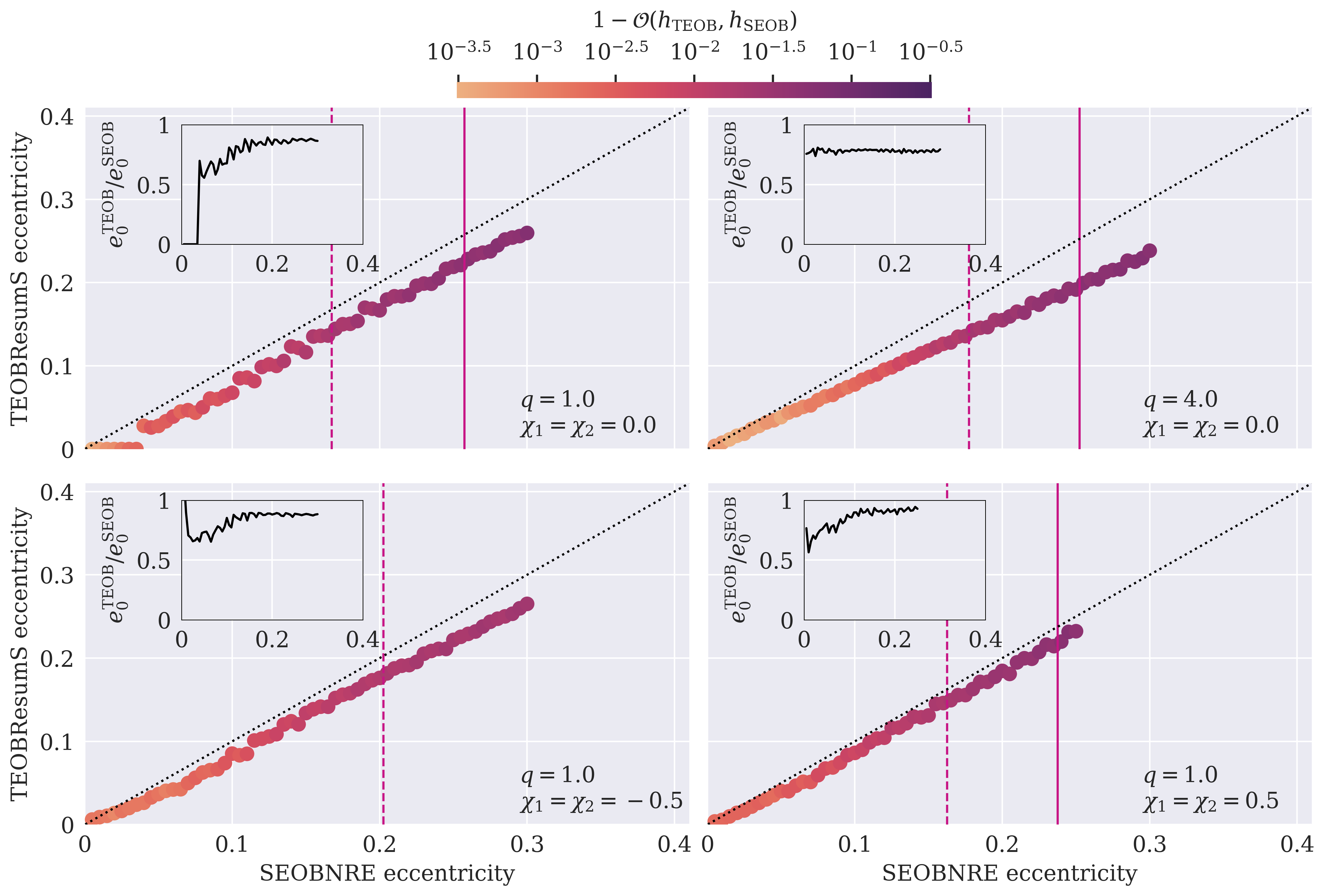}
    \caption{Effect of varying the mass ratio and component spins on the the eccentricity mapping between \texttt{TEOBResumS} and \texttt{SEOBNRE} (at 10 Hz). The total mass here is fixed at $150\,M_\odot$. The top left panel is repeated from Figure \ref{fig:ecc_conversion}. Again, the insets show the ratio of the mapped eccentricities, and the vertical lines denote the same mismatch thresholds from Figure \ref{fig:ecc_conversion}. \label{fig:ecc_conversion_params}}
\end{figure*}

In this section, we discuss the results of our analysis of \texttt{TEOBResumS} and \texttt{SEOBNRE}, using the mismatch-minimization procedure described in Section \ref{sec:method}. The mapping between the two eccentricity definitions are shown in Figure \ref{fig:ecc_conversion}, in which all eccentricities are defined at $10$~Hz. Across the parameter space we explore, we find that a smaller value for eccentricity must be passed to \texttt{TEOBResumS} compared to \texttt{SEOBNRE} in order to produce the best-matching waveform. If we represent the mapping from \texttt{SEOBNRE} eccentricity to \texttt{TEOBResumS} eccentricity by $e_0^{\rm TEOB}= k e_0^{\rm SEOB}$, then the conversion factor, $k$, typically fluctuates between $0.5$-$0.8$. In other words, for a given numerical value of \texttt{SEOBNRE} eccentricity, we must pass a $20$-$50\%$ smaller value of eccentricity to \texttt{TEOBResumS} to obtain an equivalently eccentric waveform. This is true provided the reference eccentricity passed to the waveforms is also made to be consistent, as noted in Section \ref{sec:eccwave}. 

The conversion factor is sensitive to the intrinsic parameters of the system, including both the eccentricity parameter itself as well as the masses and spins. As shown in the insets of Figure \ref{fig:ecc_conversion}, the conversion factor is stable at lower masses and high eccentricities, but as we go to lower eccentricities this factor changes, requiring larger adjustments to match the waveforms. This is likely a consequence of the models' diverging behaviour in the small-eccentricity regime (see Figure \ref{fig:mismatch_circ}). Relatedly, the eccentric versions of the two models cannot be reconciled at low but non-zero eccentricities. When the \texttt{SEOBNRE} eccentricity is sufficiently small, we find that the quasi-circular \texttt{TEOBResumS-GIOTTO} waveform yields a lower mismatch than any eccentric \texttt{TEOBResumS-DALI} waveform. Thus, Figure \ref{fig:ecc_conversion} shows that small enough \texttt{SEOBNRE} eccentricities are mapped onto zero \texttt{TEOBResumS} eccentricity. Figure \ref{fig:ecc_conversion_params} shows how the conversion varies slightly with the masses and spins, but remains between $20$-$50\%$ except in the case of larger aligned-spins, where the needed adjustment reaches as low as $10\%$.

Figure \ref{fig:ecc_conversion} shows that the minimum mismatch between the waveforms increases with larger eccentricity. We have drawn vertical lines to denote when the mismatch passes $2\%$ and $5\%$, giving a sense of the eccentricity and mass range for which our eccentricity maps yield roughly equivalent waveforms. As expected, lighter systems exceed these thresholds at lower eccentricities than heavier systems, as these system retain a larger fraction of their inspiral in-band. Since it is during the inspirals where the models are maximally different, this contributes to higher overall mismatch. For example, at detector-frame mass $100\,M_\odot$ the lowest possible mismatch between \texttt{SEOBNRE} and \texttt{TEOBResumS} is greater than $5\%$ for eccentricities $e_0^{\rm SEOB}\gtrsim 0.2$ at $10$~Hz. However, for a $250\,M_\odot$ system that mismatch never exceeds $2\%$ for $e_0^{\rm SEOB}< 0.2$.

We also observe that the relation between $e_0^{\rm SEOB}$ and $e_0^{\rm TEOB}$ appears to oscillate as a function of eccentricity for higher-mass systems, becoming more noticeable at higher masses. It is not clear yet why this occurs, but is probably related to systematic differences between the two models when measuring the eccentricity closer to merger.

\subsection{Numerical relativity analysis}

To complement our eccentricity definition study, we repeat our mismatch-minimization procedure to measure the eccentricities of numerical relativity simulations, cross-checking the results with our eccentricity maps. We analyze a set of eccentric BBH simulations provided in the SXS catalog \citep{Boyle:2019kee, Hinder:2017sxy}, which includes non-spinning and aligned-spin configurations, and mass ratios ranging from $q=1$ to 3. The waveforms are scaled to a total mass of $150\,M_\odot$, which ensures each simulation has frequency content starting below $10$~Hz, which we again integrate from in our mismatch calculations. We use only the $(\ell=2, |m|=2)$ waveforms, and we remove the initial junk radiation from the simulations.

\begin{figure}
    \epsscale{1.05}
    \plotone{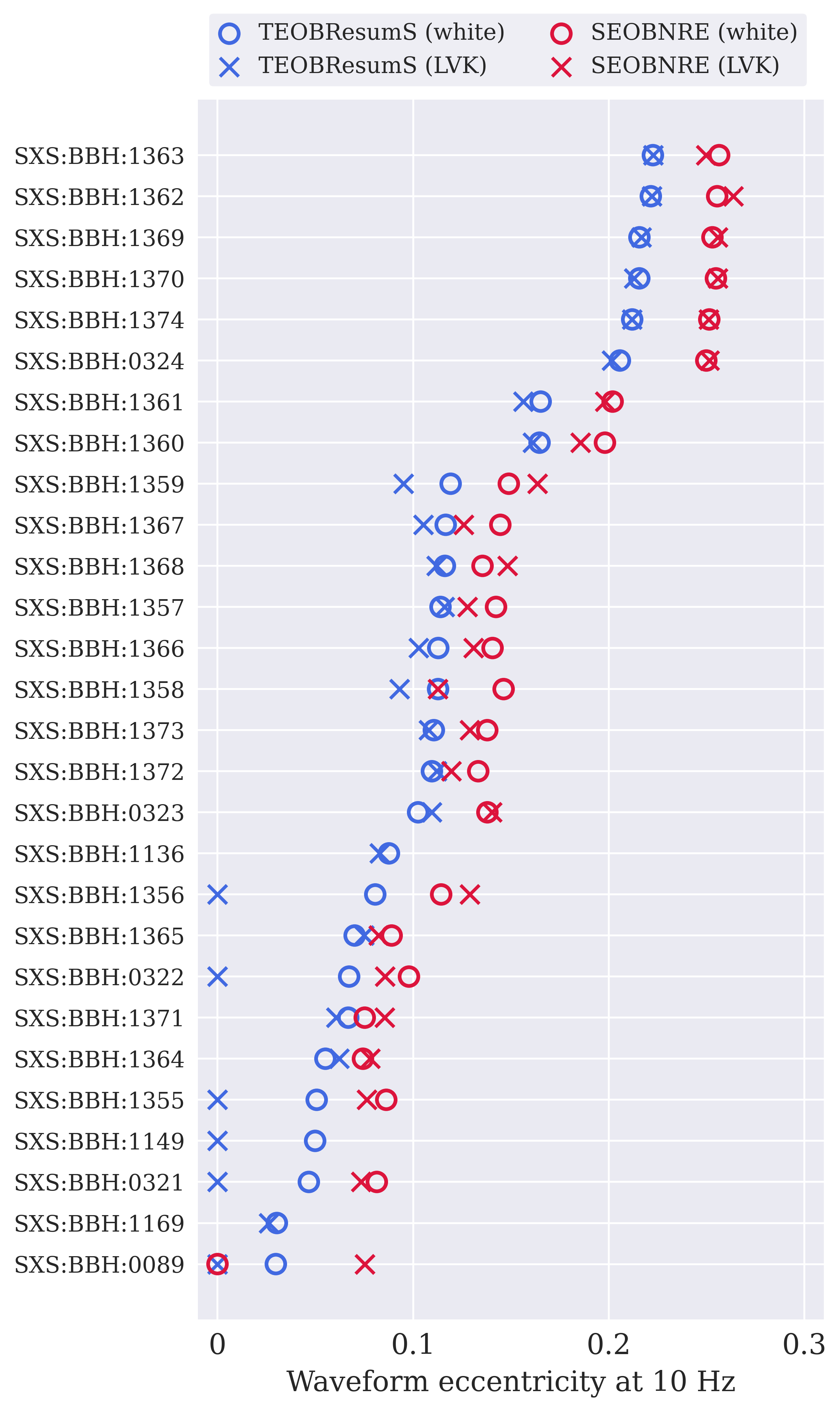}
    \caption{Eccentricities in our chosen set of numerical relativity simulations measured with \texttt{TEOBResumS} and \texttt{SEOBNRE} using our mismatch-minimization procedure. The eccentricities are evolved to a frequency of $10$~Hz. The simulations were scaled to a total mass of $150\,M_\odot$ in the detector frame, and the mismatches are computed with respect to either white (circles) or LVK (crosses) noise, integrating from $10$~Hz. Note that three simulations (\texttt{SXS:BBH:1136}, \texttt{SXS:BBH:1149}, \texttt{SXS:BBH:1169}) lack an \texttt{SEOBNRE} measurement due to having spins outside the allowed range of this model, $|\chi_{1,2}|< 0.6$. \label{fig:mismatch_nr}}
\end{figure}

The measured eccentricities are shown in Figure \ref{fig:mismatch_nr}, where we report results assuming both white and LVK noise. These results are consistent with our eccentricity maps, with \texttt{TEOBResumS} consistently measuring a smaller eccentricity than \texttt{SEOBNRE} to compensate for its larger simulated eccentricity. At LVK sensitivity, we find that \texttt{TEOBResumS-DALI} cannot yield as low of a mismatch as \texttt{TEOBResumS-GIOTTO} for a subset of the simulations, indicated by a measured eccentricity of zero. This is due to a combination of factors, namely the differing behaviours of the two models at small eccentricities, and because the detector PSDs down-weight the inspiral frequencies (where the eccentricity is largest) relative to the merger frequencies.

\subsection{Limitations}

\texttt{TEOBResumS} and \texttt{SEOBNRE} possess multiple differences aside from their different definitions of eccentricity. Consequently, there will be potentially measurable differences between the waveforms that cannot be completely eliminated by varying their parameters. Furthermore, our procedure does not uniquely fit for the different eccentricity definitions. Systematic differences between the models related to their treatment of radiation-reaction, the conservative dynamics (via the definition of the Hamiltonian), and their numerical relativity calibration will inevitably affect the best-fitting parameters to some degree. We examine the utility of our eccentricity maps by comparing the optimal signal-to-noise ratio (SNR) of the residuals between the mapped \texttt{TEOBResumS} and \texttt{SEOBNRE} waveforms with their minimum mismatch. This tells us whether these difference are detectable at LVK sensitivity, which would degrade the usefulness of our eccentricity maps.

For a given PSD, the optimal SNR is defined as the noise-weighted inner product of the waveform with itself,
\begin{equation}
    \rho_0 = \sqrt{\langle \delta h|\delta h\rangle}\,,
\end{equation}
where $\delta h = h_{\rm TEOB} - h_{\rm SEOB}$ is the residual strain after aligning the waveforms in coalescence time and phase. As a general rule, this difference is potentially detectable if \citep{Flanagan:1997kp, Lower:2018seu}
\begin{equation} \label{eq:detectcrit}
    1-\mathcal{O} \gtrsim \rho_0^{-2}\,,
\end{equation}
where we average over sky location, polarization angle, and orbital inclination. Since the mismatch calculations used to construct the eccentricity maps assume white noise PSDs, we recompute these mismatches under LVK sensitivity to facilitate comparison with the SNRs. If the above condition is not met, i.e.~that $1-\mathcal{O}<\rho_0^{-2}$, then no differences between the waveforms can be detected \citep{Purrer:2019jcp}.\footnote{Note that \citet{Chatziioannou:2017tdw} and \citet{Purrer:2019jcp} argue that the right-hand-side of Equation \ref{eq:detectcrit} should contain a factor of $D-1$, where $D$ is the number of waveform parameters. Including this factor would increase the mismatch threshold required to distinguish the waveforms, expanding the region of parameter space where the waveforms cannot be distinguished. Equation \ref{eq:detectcrit} thus gives a more conservative range for where our eccentricity maps are most useful.} 

\begin{figure}
    \epsscale{1.15}
    \plotone{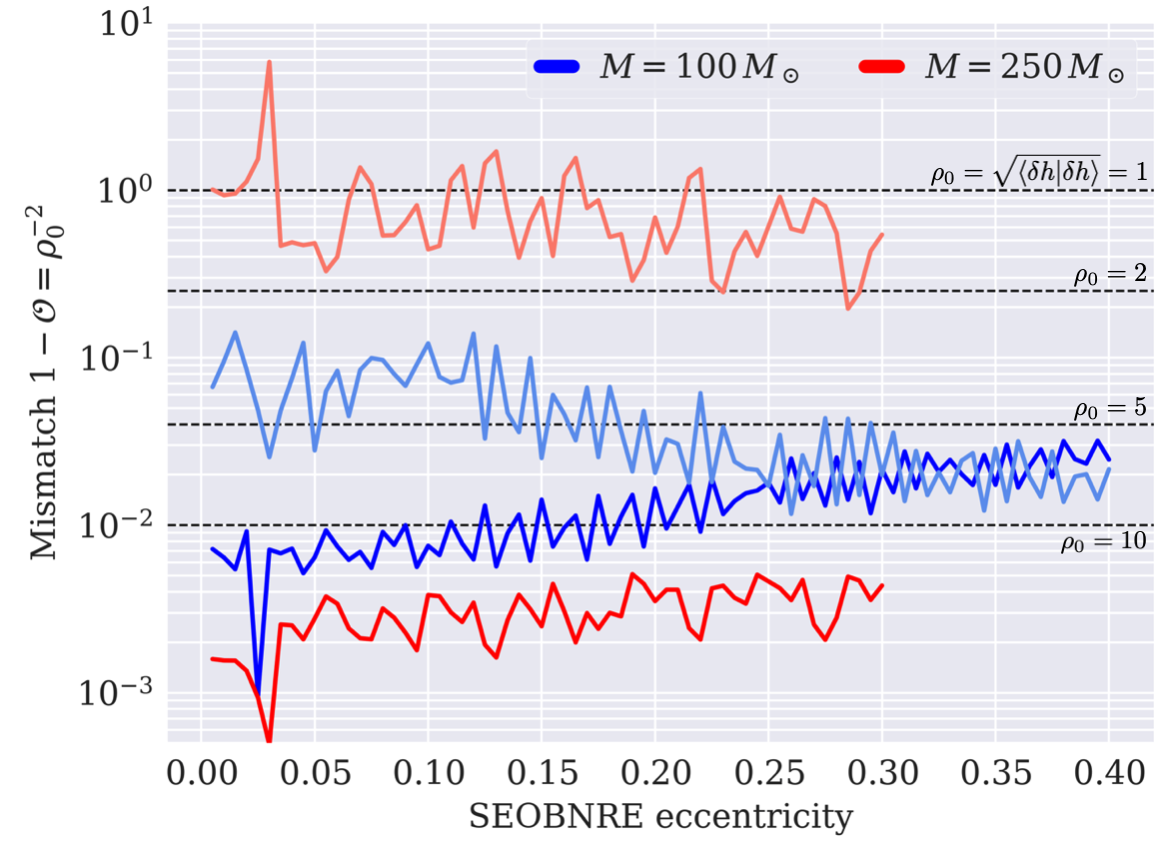}
    \caption{Detectability of the residual differences between \texttt{SEOBNRE} and \texttt{TEOBResumS} after minimizing their mismatch. The dark blue (red) curve shows the minimum mismatch between the waveforms, assuming LVK design sensitivity, as a function of \texttt{SEOBNRE} eccentricity at 10 Hz for a $M=100\,M_\odot$ ($M=250\,M_\odot$) BBH at a distance of 1 Gpc (5 Gpc). The light red/blue curves show the inverse squared optimal SNR of the difference, $\delta h = h_{\rm TEOB} - h_{\rm SEOB}$, between the mismatch-minimized waveforms, again computed with LVK sensitivity. For reference, we have drawn horizontal dashed lines for certain values of optimal SNR. The differences are detectable wherever the darker-shaded curve crosses above the lighter curve of the same base colour. The differences between the waveforms for $100\,M_\odot$ at 1 Gpc become detectable above roughly $e_0^{\rm SEOB}=0.25$, whereas they are never detectable for the $M=250\,M_\odot$ system at 5 Gpc in the parameter range we explored. \label{fig:mismatch_snr}}
\end{figure}

Figure \ref{fig:mismatch_snr} shows the results of our mismatch-SNR comparison. We consider two classes of sources in our analysis: a $100\,M_\odot$ BBH (detector frame) located at $1$~Gpc, and a $250\,M_\odot$ BBH located at 5 Gpc. We find it is unlikely that we would be able to detect the residual difference across much of the parameter space, especially in distant, high-mass sources with only a handful of cycles in-band such as GW190521. For such systems, the SNR of the residuals registers between $1$-$2$ at LVK sensitivity. Our method will be less effective for less massive, highly eccentric systems, where we see a greater build-up in SNR from $\delta h$ due to the substantially longer observable inspiral. For our specific example of a $100\,M_\odot$ BBH shown in Figure \ref{fig:mismatch_snr}, the residual SNR reaches the $5$-$10$ range for eccentricities $e_0^{\rm SEOB} \gtrsim 0.25$ at $10$~Hz, where it starts to be comparable to the mismatch. We note, however, that this eccentricity range is above what was considered by \citet{Romero-Shaw:2019itr, Romero-Shaw:2020thy, Romero-Shaw:2021ual, Romero-Shaw:2022xko}, which probed \texttt{SEOBNRE} eccentricities up to only $e_0^{\rm SEOB}=0.2$. If the system was located further away, then we would be even less likely to detect any difference between the mapped waveforms.

\section{Summary} \label{sec:summary}

The importance of orbital eccentricity as a probe of dynamical BBH formation underscores the need to have an in-depth understanding of how eccentricity is defined by different waveform models. These models can also differ in the definition of the reference frequency and initial mean anomaly, further complicating the interpretation of our measurements.

In this work, we perform various studies aimed at reconciling the eccentricity definitions used by two eccentric EOB waveform models, \texttt{TEOBREsumS} and \texttt{SEOBNRE}. We find the optimal values of the waveform eccentricities and reference frequencies which minimize the mismatch between the models, and then evolve the eccentricities to a common frequency of $10$~Hz. Our analysis shows that \texttt{TEOBResumS} simulates a larger eccentricity for the same numerical input at \texttt{SEOBNRE}, which must be adjusted downwards to match the eccentricity of \texttt{SEOBNRE}. However, we also find that our method does not work to resolve very small eccentricities due to the behaviours of the two models at small eccentricities. In addition, we highlight other waveform differences that must be accounted for when comparing parameter estimation results, mainly that the waveforms initialize the system from opposite orbital positions, and the use of inconsistent reference frequencies. 

We supplement this work with a comparison to a set of eccentric numerical relativity simulations, showing that the eccentricities measured by \texttt{TEOBResumS} and \texttt{SEOBNRE} via our mismatch procedure are consistent with our established maps. Additional work, however, is still required to fully understand why these two models yield different eccentricity measurements, which we discuss briefly in Appendix \ref{sec:appendix_gw}.

Lastly, we examine the robustness of our method in terms of producing a faithful conversion between the two models. For low masses and high eccentricities, our mismatch minimization returns waveforms with residual mismatches that are potentially detectable at LVK sensitivity. For heavier and more distant systems like GW190521, we find that these differences are likely not detectable. The definition of eccentricity could potentially be disentangled from other features of the waveform by measuring the eccentricity directly from its frequency evolution. The exploration of a standardized eccentricity parameter that could be used to compare measurements across different models is currently underway \citep{Shaikh2022, Bonino2022}.

\begin{acknowledgments}

We thank Alessandro Nagar, Rossella Gamba, Piero Rettegno, and Zhoujian Cao for answering our queries about their respective waveform models. We thank Vijay Varma and Harald Pfeiffer for helpful discussions concerning eccentricity definitions. We also thank Harald Pfeiffer for providing comments on this manuscript. This material is based upon work supported by NSF's LIGO Laboratory which is a major facility fully funded by the National Science Foundation. We are grateful for computational resources provided by the LIGO Laboratory and supported by National Science Foundation Grants PHY-0757058 and PHY-0823459.
A.M.K. and J.M. acknowledge funding support from the Natural Sciences and Engineering Research Council of Canada (NSERC) through the Discovery Grants program. 
A.M.K. was also supported by the John I. Watters Research Fellowship. 
I.M.R.-S, P.D.L, and E.T. acknowledge support from the Australian Research Council (ARC) Centre of Excellence CE170100004.
I.M.R.-S. also acknowledges support received from the Herchel Smith Postdoctoral Fellowship Fund.

\end{acknowledgments}

%


\software{\texttt{PyCBC} \citep{2019PASP..131b4503B},
          \texttt{LALSuite} \citep{lalsuite},
          \texttt{SciPy} \citep{2020SciPy-NMeth},
          \texttt{Bilby} \citep{2019ApJS..241...27A, 2020MNRAS.499.3295R},
          \texttt{Matplotlib} \citep{Hunter:2007}
          }



\appendix

\section{Eccentricity maps assuming LVK sensitivity} \label{sec:appendix_lvk}

\begin{figure*}
    \epsscale{1.05}
    \plotone{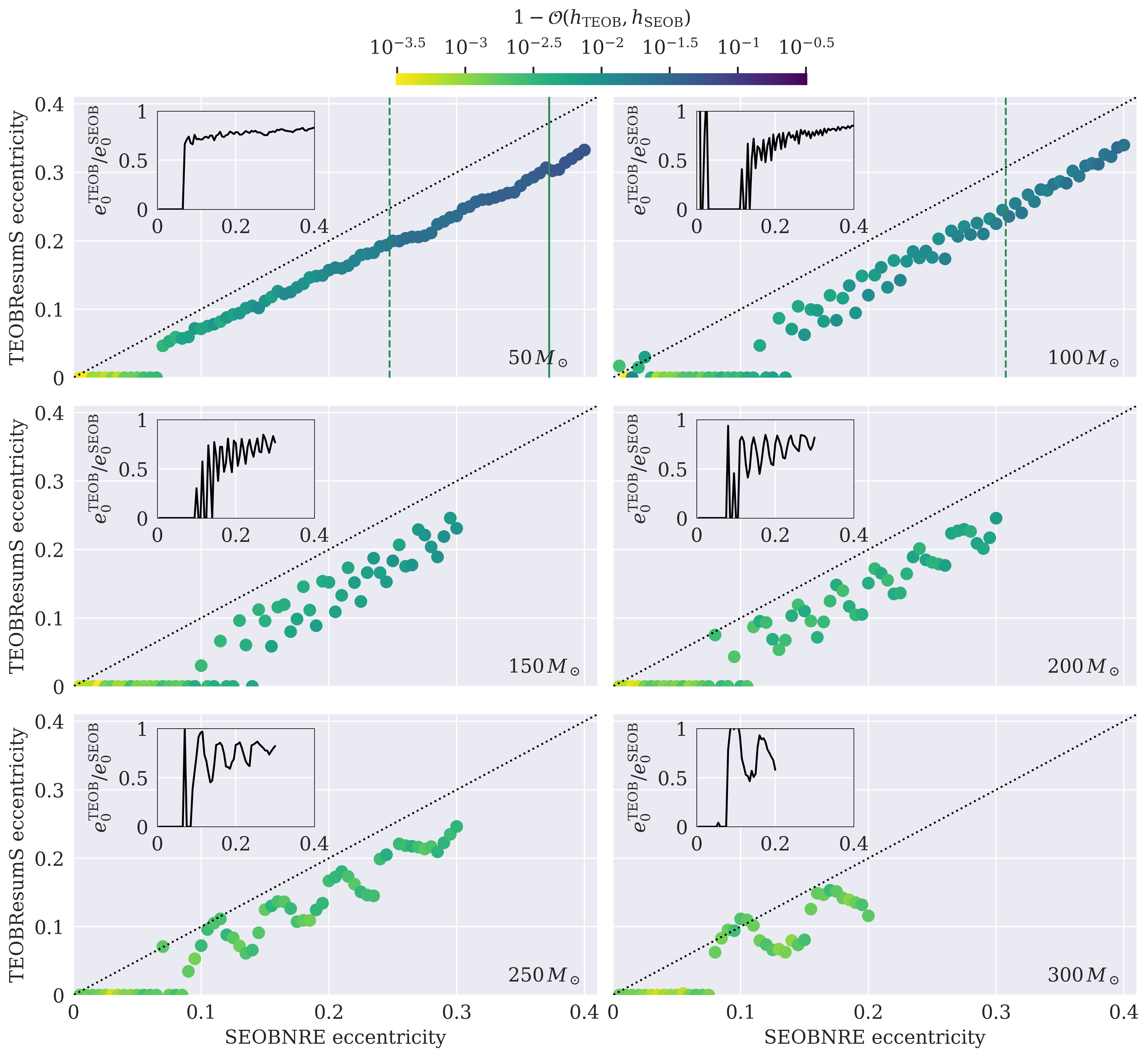}
    \caption{Eccentricity maps calculated with LVK sensitivity instead of white noise. This plot shows that the same information as in Figure \ref{fig:ecc_conversion}, except we replace the flat PSDs with detector PSDs in the calculation of the mismatch. \label{fig:ecc_conversion_lvk}}
\end{figure*}

Figure \ref{fig:ecc_conversion_lvk} shows how our eccentricity maps appear when using LVK sensitivity to calculate the waveform mismatch. Specifically, we used the Advanced LIGO zero-detuned high-power noise curve provided in \texttt{LALSimulation} \citep{lalsuite}. The relation between the two eccentricities is noticeably less smooth, as shown in Figure \ref{fig:ecc_conversion_lvk}. Compared to our white noise results, the eccentricity of \texttt{SEOBNRE} must be much higher before the lowest-mismatch \texttt{TEOBResumS} waveform is not quasi-circular. This occurs because the LVK sensitivity curve weighs higher frequencies near merger more strongly than deeper in the inspiral, and so the system must have larger eccentricity closer to merger before it can be detected with \texttt{TEOBResumS}.

\section{Eccentricity constraints on GW events} \label{sec:appendix_gw}

In Figure \ref{fig:gwconstraints}, we show eccentricity constraints obtained with \texttt{SEOBNRE} and \texttt{TEOBResumS} for ten selected GW events detected by the LVK. These events are chosen because, of all events in GWTC-2, they have the highest support for the eccentric-waveform hypothesis when analysed with \texttt{SEOBNRE} \citep{Romero-Shaw:2021ual}. We employ the Bayesian inference package \texttt{Bilby} \citep{2019ApJS..241...27A, 2020MNRAS.499.3295R} to perform parameter estimation on these events. We reuse posterior samples for \texttt{SEOBNRE} that were first presented in previous work \citep[][]{Romero-Shaw:2021ual}, which were obtained by first sampling the posterior with a quasi-circular waveform model, \texttt{IMRPhenomD}, and then reweighting with \texttt{SEOBNRE} (see \citet{Romero-Shaw:2019itr, Payne:2019wmy} for a full description of the likelihood reweighting process). Posteriors for \texttt{TEOBResumS} are obtained by using this model directly in parameter estimation, without any reweighting, but with a loosened error tolerance as demonstrated in \citet{OShea:2021ugg}. For \texttt{TEOBResumS}, we use a log-uniform prior on eccentricity in the range $10^{-4} \leq e_{10} \leq 0.3$. 

\subsection{Discussion of GW190521}

\begin{figure*}
    \epsscale{0.65}
    \plotone{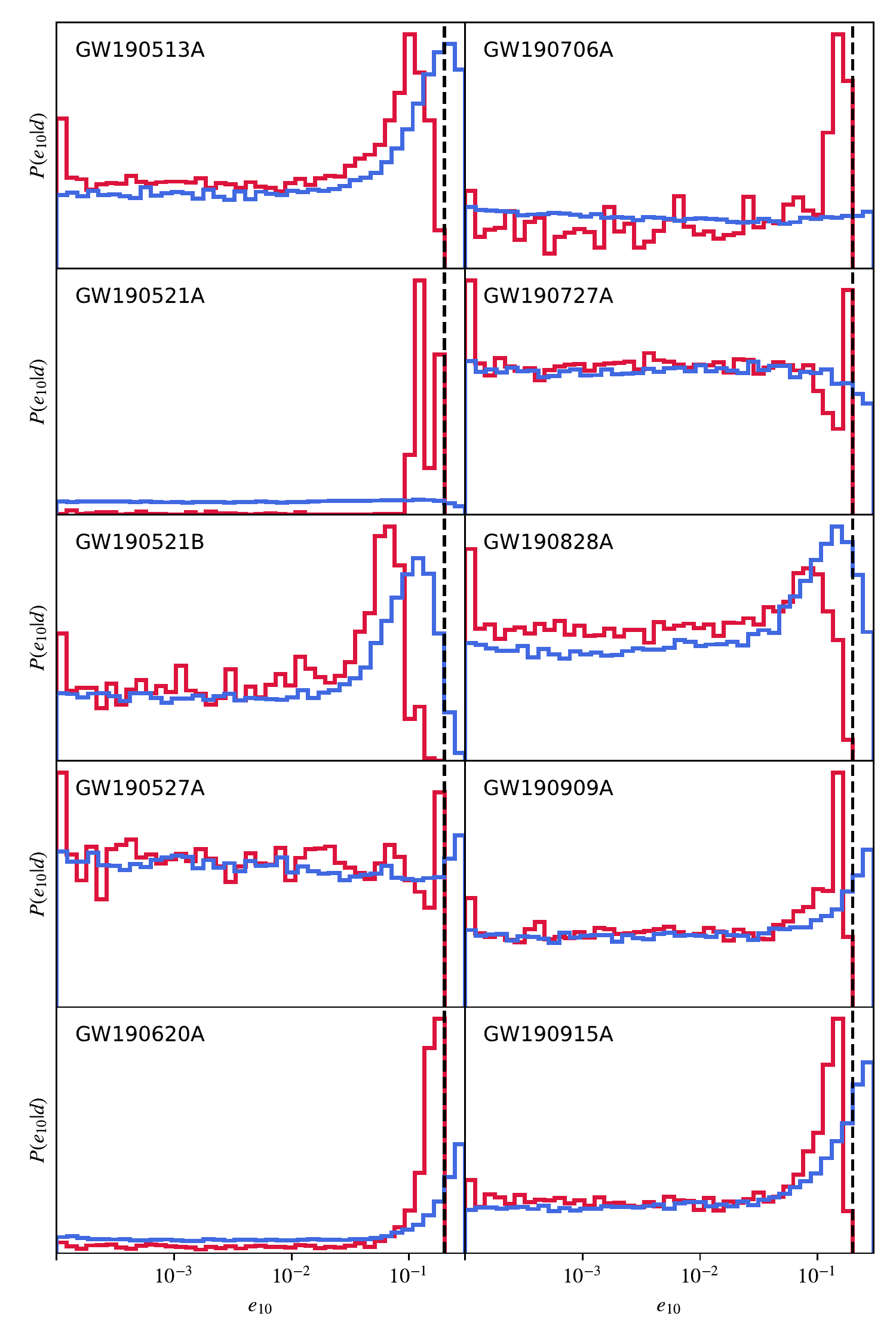}
    \caption{Marginal posterior probability distributions for eccentricity on ten selected GW events, obtained with either \texttt{SEOBNRE} (red) or \texttt{TEOBResumS} (blue). Note that GW190909 has since been downgraded to a non-astrophysical trigger \citep{LIGOScientific:2021usb}. In many cases the peaks in the posterior distributions rail against the upper prior limit. The upper prior limit used for \texttt{TEOBResumS} is at the right edges of the plots. The upper prior limit of \texttt{SEOBNRE} is shown by the dashed vertical lines. The upper prior limit of \texttt{TEOBResumS} is $e_{0}^{\rm TEOB} = 0.3$ at 10 Hz. \label{fig:gwconstraints}}
\end{figure*}

Figure \ref{fig:gwconstraints} shows that we infer different eccentricity posterior distributions between \texttt{SEOBNRE} and \texttt{TEOBResumS} for these ten events, with the latter yielding either an uninformative (uniform) distribution, or a distribution that is peaked slightly above the \texttt{SEOBNRE} posterior. Despite being peaked away from zero eccentricity, most of the events appear consistent with quasi-circularity, as their posteriors have support over the full prior range. The exception to this is GW190521, which shows more evidence for the eccentric hypothesis when analyzed with \texttt{SEOBNRE} \citep{Romero-Shaw:2020thy}, with little support for $e_0^{\rm SEOB}<0.1$ at 10 Hz. However, when analyzed with \texttt{TEOBResumS}, the posterior is uninformative.
This discrepancy is curious.

We suggest a few possibilities as to why we infer different posteriors for GW190521. One possibility is that the bulk of the \texttt{TEOBResumS} posterior support is located above the upper limit of the prior, causing us to miss the true peak in the eccentricity posterior. This hypothesis is supported by the fact that the peak of the \texttt{SEOBNRE} posterior is very close to the prior limit. Note, however, that our eccentricity mapping study shows that a given \texttt{SEOBNRE} eccentricity will translate into a smaller numerical value of \texttt{TEOBResumS} eccentricity after evolving to consistent reference frequencies. Thus, if the \texttt{TEOBResumS} distribution is indeed peaked at a higher eccentricity, then it is not measuring the same physical eccentricity as \texttt{SEOBNRE}. An alternative explanation is that the \texttt{TEOBResumS} posterior is consistent with quasi-circularity, as Figure \ref{fig:ecc_conversion_lvk} shows that systems with $e_0^{\rm SEOB}\lesssim 0.15$ at 10 Hz can have a best-matching \texttt{TEOBResumS} waveform with $e_0^{\rm TEOB}=0$ assuming LVK noise. This can occur because the sensitivity curves down-weight the lower inspiral frequencies relative to the merger, making it harder to detect any eccentricity. 
A third possibility is that the event is not eccentric at all, but it is precessing, and the precessing waveform is best fit by \texttt{SEOBNRE} with $e_0^{\rm SEOB}>0$ while the \texttt{TEOBResumS} waveform can be fit using $e_0^{\rm TEOB} \approx 0$.

The inconsistent results between \texttt{SEOBNRE} and \texttt{TEOBResumS} are not yet understood. Our eccentricity maps are a step forward in this regard, as it partially disentangles the definition difference from other systematics. However, further work is needed to determine how one waveform could measure significant eccentricity while the other measures none. This will likely require injection studies in which one adds an eccentric (and/or precessing) waveform to realistic detector noise and attempts to recover the signal with a different eccentric model by running full parameter estimation, instead of calculating mismatches as done in this work.


Comparing our results to those of \citet{Iglesias2022}, we find qualitatively consistent eccentricity posterior distributions, when differences in prior shape, prior volume, reference frequency, analysis technique, and waveform model are considered.
In particular, we note that their use of a higher reference frequency both reduces the eccentricity and reduces the evidence for eccentricity, thereby disfavouring the eccentric hypothesis.


\bibliography{biblio}{}
\bibliographystyle{aasjournal}



\end{document}